\documentclass[twoside,7pt]{amsart}
\usepackage{amsmath,amssymb,mathrsfs}
\usepackage[hypertex,linkcolor=red]{hyperref}
\def\d{\operatorname{d}}\def\<{\langle}\def\>{\rangle}
\def\Tr{\operatorname{Tr}}\def\:{\hbox{\bf :}}
\def\Cmplx{\mathbb C}

\def\map#1{{\mathcal{#1}}}
\def\set#1{{\sf #1}}\def\alg#1{{\mathcal #1}}\def\spc#1{{\mathscr{#1}}} 
\def\Bnd#1{\alg{B(\spc{#1})}}\def\Bndd#1{\alg{B({#1})}}\def\Conv{\set{C}}

\def\vec#1{{\boldsymbol{#1}}}
\def\sH{\spc{H}}\def\sM{\spc{M}}\def\sR{\spc{R}}

\def\grp#1{{\mathbf #1}}

\def\rnk{\operatorname{rank}}
\def\Rng{\set{Rng}}\def\Ker{\set{Ker}}\def\Supp{\set{Supp}}
\def\Klm{{\mathfrak X}}\def\dim{\operatorname{dim}}
\def\Span{\set{Span}}

\def\gG{\grp{G}}\def\gH{\grp{H}}
\def\qed{$\,\blacksquare$\par}

\def\dag{\dagger}
\def\Det{\operatorname{Det}}
\newtheorem{lemma}{Lemma}
\newtheorem{proposition}{Proposition}
\newtheorem{corollary}{Corollary}
\newtheorem{theorem}{Theorem}
\def\Proof{\medskip\par\noindent{\bf Proof. }}
\def\implies{\Longrightarrow}
\begin{document}
\title{Extremal covariant POVM's}
\author{Giulio Chiribella}
\email{chiribella@unipv.it}
\address{{\em QUIT} Group, http://www.qubit.it, Istituto Nazionale di Fisica della Materia,
Unit\`a di Pavia, Dipartimento di Fisica "A. Volta", via Bassi 6,
I-27100 Pavia, Italy}
\author{Giacomo Mauro D'Ariano}
\email{dariano@unipv.it}
\address{{\em QUIT} Group, http://www.qubit.it, Istituto Nazionale di Fisica della Materia,
Unit\`a di Pavia, Dipartimento di Fisica "A. Volta", via Bassi 6,
I-27100 Pavia, Italy, and \\ Department of Electrical and Computer
Engineering, Northwestern University, Evanston, IL  60208}
\date{\today}
\maketitle
\begin{abstract}
We consider the convex set of positive operator valued measures (POVM) which are covariant under a
finite dimensional unitary projective representation of a group. We derive a general
characterization for the extremal points, and provide bounds for the ranks of the corresponding POVM
densities, also relating extremality to uniqueness and stability of optimized measurements.
Examples of  applications are given. 
\end{abstract}     
\section{introduction}
An essential step in the design of the new quantum information technology\cite{Nielsen2000} is to
asses the ultimate precision limits achievable by quantum measurements in extracting information
from physical systems. For example, the security analysis of a quantum cryptographic 
protocol\cite{gisirev} is based on the evaluation of the limits posed in principle by the quantum
laws to any possible eavesdropping strategy. A general method to establish such limits is to
optimize a quantum measurement according to a suitable criterion, and this is the general objective
of the so-called {\em quantum estimation theory}\cite{helstrom, holevo}. Different criteria can
be adopted for optimizing the measurement, the choice of a particular one depending on the
particular problem at hand. Moreover, many different optimization problems often share the same
form, e. g. they resort to the maximization of a concave function on the set of the possible
measurements. We remind that measurements form a convex set, the convex combination corresponding 
to the random choice between two different apparatuses. Since a concave function attains its
maximum in an extremal point, it is clear that the optimization problem is strictly connected to the
problem of characterizing the extremal points of the convex set. 

The quantum measurements interesting in most applications are {\em covariant}\cite{holevo}
with respect to a group of physical transformations. In a purely statistical description of a
quantum measurement in terms of the outcome probability only---i. e. without considering the 
state-reduction---the measurement is completely described by a positive operator valued measure 
(POVM) on its probability space. In terms of POVM's, "group-covariant" means that there is an action of the
transformation group on the probability space which maps events into events, in such 
a way that when the measured system is transformed according to a group transformation, the
probability of a given event becomes the probability of the transformed event. Such scenario
naturally occurs in the estimation of an unknown group transformation performed on a known input
state, e. g. in the estimation of the unknown unitary transformation\cite{entang_meas,ajv}, in the
measurement of a phase-shift in the radiation field \cite{holevo, DMS}, or in the estimation of
rotations on a system of spins \cite{refframe}. A first technique for characterizing extremal
covariant POVM's and quantum operations has been presented in Ref. \cite{extPOVMandQO}
inspired by the method for characterizing extremal correlation matrices of Ref. \cite{LiTam}, in particular,
classification of extremal POVM's has been presented for the case of trivial stability group, i. e.
when the only transformation which leaves the input state unchanged is the identity. Here we solve
the characterization problem for extremal covariant POVM's in the general case of nontrivial
stability group, providing a simple criterion for extremality in Theorem \ref{supporti} in terms of
minimality of the support of the {\em seed} of the POVM, presenting iff conditions for extremality
in Theorem \ref{th:iff}, and providing bounds for the rank of extremal POVM's (in the following we
will define the rank of a POVM as the rank of its respective density: see Eq. (\ref{density}) for
its definition).
We show that, contrarily to the usual credo, the optimal covariant POVM can have rank larger
than one. Indeed, there are group representations for which covariant POVM cannot
have unit rank, since this would violate a general bound for the rank of the POVM in
relation to dimensions and multiplicity of the invariant subspaces of the group.
In the present paper we adopt the maximum likelihood optimality criterion, which, however, as we
will show, is formally equivalent to the solution of  the optimization problem in a very large class of
optimality criteria. Other issues of practical interest that we address are the uniqueness and the
stability of the optimal covariant POVM. The whole derivation is given for finite dimensional
Hilbert spaces: as we will show in a simple example, it can be generalized to infinite dimensions,
however, at the price of  making the theory much more technical. 
\par The paper is organized as follows. After introducing covariant POVM's and their convex
structure in Section \ref{ConvStruct}, the main group theoretical tools that will be used for the  
characterization of covariant POVM's  are presented in Section \ref{GrpTools}. In Section
\ref{ExtrPovms} we give a characterization of extremal covariant POVM's in finite dimension with
a general stability group, deriving an algebraic extremality criterion, along with a general bound for the rank of the extremal POVM's in
terms of the dimensions of the invariant subspaces of the group and of the stability subgroup.
Properties of extremal POVM's in relation with optimization problems are analyzed in 
Section \ref{OptExtr}, where also the issues of uniqueness and stability of the optimal covariant
POVM's are addressed. Finally, examples of application of the theory to estimation of rotation,
state, phase-shift, etc. are given in Section \ref{s:examples}, providing extremal POVM's with a non
trivial stability group and giving examples of optimization problems with solution consisting of
extremal POVM with rank greater than one.
\section{Convex structure of covariant POVM's}\label{ConvStruct}
The general description of the statistics of a measurement is given in
terms of a probability space $\Klm$---the set of all possible
measurement {\em outcomes}---equipped with a $\sigma-$algebra
$\sigma(\Klm)$ of subsets $\set{B}\subseteq\Klm$ and with a probability
measure $p$ on $\sigma(\Klm)$. Each subset $\set{B}\in \sigma(\Klm)$
describes the event "the outcome $x$ belongs to $\set{B}$" and the
statistics of the measurement is fully specified by the probability
measure $p$, which associates to any event $\set{B}$ its probability
$p(\set{B})$.

In quantum mechanics the probability $p(\set{B})$ is given by the Born rule
\begin{equation}
p(\set{B}) \doteq \Tr[\rho P(\set{B})]
\end{equation}
where $\rho$ is a density operator (i.e. a positive semidefinite operator with unit trace) on the Hilbert space $\sH$ of the
measured system, representing its state, whereas $P$ is the POVM of
the apparatus, giving the probability measure $p$ for every given
state $\rho$ of the quantum system. Mathematically a POVM $P:
\sigma(\Klm) \to \Bnd H$ is a \emph{positive operator valued measure}
on $\sigma(\Klm)$, namely it satisfies the following defining
properties
\begin{eqnarray}
&&0 \leq P(\set{B}) \leq I \qquad \forall \set{B}\in \sigma(\Klm)\\
&&P(\cup_{i=1}^{\infty} \set{B}_i)= \sum_{i=1}^{\infty} P(\set{B}_i)\quad \forall
\{\set{B}_i\}~~\text{disjoint}\\ &&P(\Klm)=I.
\end{eqnarray}   
Notice that the set of POVM's for $\sigma(\Klm)$ is a convex set, namely, if $P_1$ and $P_2$ are
POVM's for $\sigma(\Klm)$, then also $\lambda P_1+(1-\lambda)P_2$ is a POVM for 
$\sigma(\Klm)$ for any $0\leq\lambda\leq 1$. The measurement described by the POVM $\lambda
P_1+(1-\lambda)P_2$  corresponds to randomly choosing between two different measuring apparatuses
described by the POVM's  $P_1$ and $P_2$ respectively. The extremal points of such convex set of
POVM's---the socalled \emph{extremal POVM's}---correspond to measurements that cannot result
from a random choice between different measuring apparatuses.

In the following we will focus attention to the case of probability space $\Klm$ given by the
quotient $\gG/\gG_0$ of a compact Lie group $\gG$ with respect to a subgroup
$\gG_0$. Physically, this situation arises when the POVM is designed to estimate a state
of the group-orbit $\{U_g \rho U_g^{\dag}~|~ g \in \gG\}$ of a given state $\rho$, with the group
$\gG$ acting on the Hilbert space $\sH$ of a quantum system via the unitary projective
representation $\set{R}(\gG)\doteq\{U_g
~|~ g \in \gG\}$. In such case, in fact, the probability space of the POVM is exactly $\Klm
=\gG/\gG_\rho$, and $\gG_0=\{h\in \gG~|~U_h \rho U_h^{\dag}=\rho\}$ is the stability 
group of  $\rho$, whence the points of the orbit are in one to one 
correspondence with the elements of $\Klm=\gG/\gG_0$. Notice that in the following the fact that
the representation is projective is inconsequential, whence there will be no need of reminding it.

An important class of measurements with $\Klm = \gG /\gG_0$  is described by the \emph{covariant}
POVM's \cite{holevo}, namely those POVM's which enjoy the property
\begin{equation}
P(\set{gB})= U_g P(\set{B}) U_g^\dag \qquad \forall \set{B}\in \sigma(\Klm),~ \forall g\in\gG,
\end{equation}
where $g\set{B}\doteq \{gx ~|~ x\in \set{B}\}$.  Any POVM $P$ in this class is absolutely 
continuous with respect to the measure $\d x$  induced on $\Klm$ by the normalized Haar
measure $\d g$ on the group $\gG$, and admits an operator density $M$, namely
\begin{equation}
M:~ \Klm \to \Bnd{H},\qquad P(\set{B}) =\int_\set{B} \d x\, M(x).\label{density}
\end{equation}
For a covariant POVM, the operator density has the form \cite{holevo}
\begin{equation}\label{COVdens}
M(x)=U_{g(x)}\Xi U_{g(x)}^{\dag},
\end{equation}
where $g(x) \in \gG$ is any element in the equivalence class $x\in\Klm=\gG/\gG_0$, and $\Xi$ is an
Hermitian operator satisfying the constraints
\begin{eqnarray}
\label{XIposnorm} &&\Xi \geq 0,\qquad \int_{\gG} \d g~ U_g \Xi U_g^{\dag}~~ =I \\
\label{XIcomm} &&\left[\Xi, U_h \right] =0 \quad \forall h \in\gG_0. 
\end{eqnarray} 
The operator $\Xi$ is usually referred to as the \emph{seed} of the covariant POVM\cite{note}.

Notice that the constraints (\ref{XIposnorm}) are needed for positivity and normalization of the
probability density, whereas identity (\ref{XIcomm}) guarantees that $M(x)= U_{g(x)} \Xi
U_{g(x)}^{\dag}$ does not depend on the particular element $g(x)$ in the equivalence class $x$.  It
is easy to see that the constraints (\ref{XIposnorm}) and (\ref{XIcomm}) still define a convex set
$\Conv$, namely, 
for any $\Xi_1, \Xi_2 \in \Conv$ and for any $0\leq\lambda\leq 1$ one has $\lambda\Xi_1+ 
(1-\lambda) \Xi_2 \in \Conv$. Precisely, the convex set $\Conv$ is the intersection of the cone of
positive semidefinite operators with the two affine hyperplanes given by identity (\ref{XIcomm}) and by the
normalization condition in Eq. (\ref{XIposnorm}). Since a covariant POVM is completely specified by
its seed $\Xi$ as in Eq. (\ref{COVdens}), the classification of the the extremal covariant POVM's 
resorts to the classification of the extremal points in the convex set $\Conv$.
\section{Group theoretic tools}\label{GrpTools}
Let $\gG$ be a compact Lie group, with invariant Haar measure 
$\d g$ normalized as $\int_{\gG} \d g=1$, and consider a unitary representation
$\set{R}(\gG)=\{U_g~|~ g\in\gG\}$ on a finite dimensional Hilbert space $\sH$.
Then $\sH$ is decomposed as direct sum of orthogonal irreducible subspaces as follows
\begin{equation}
\sH = \bigoplus_{\mu \in S} \bigoplus_{i=1}^{m_{\mu}}\sH_i^{(\mu)},\label{decomp1}
\end{equation}
$\set{S}$ denoting the collection of equivalence classes of irreducible components of the
representation, the classes being labeled by the Greek index $\mu$, whereas the Latin index $i$
numbers equivalent representations in the same class. Let $T_{ij}^{(\mu)}: \sH_j^{(\mu)} \to
\sH_i^{(\mu)}$ denote invariant isomorphisms connecting the irreducible representations of the
equivalence class $\mu$ of dimension $d_{\mu}$, namely for any $i,j=1, \dots, m_{\mu}$ 
$T_{ij}^{(\mu)}: \sH_j^{(\mu)} \to \sH_i^{(\mu)}$ is an invertible operator satisfying the identity
\begin{equation}
U_g T_{ij}^{(\mu)} U_g^{\dag}=T_{ij}^{(\mu)}, \quad \forall g \in \gG.
\end{equation}
Consistently with this notation $T^{(\mu)}_{ii}$ will denote the projection operator on
$\sH_i^{(\mu)}$.  Since all subspaces $\sH_i^{(\mu)}$ are isomorphic, we can equivalently write 
\begin{equation}
\bigoplus_{i=1}^{m_{\mu}} \sH_i^{(\mu)} \equiv \sH_{\mu} \bigotimes \sM_{\mu},
\end{equation}
where  $\sH_{\mu}$ denotes the \emph{representation space}, i.e. an abstract $d_{\mu}$-dimensional subspace where a representation
of the class $\mu$ acts, while  $\sM_{\mu}$ denotes the \emph{multiplicity space}, i.e. a $m_{\mu}$-dimensional space which is unaffected
by the action of the group. 
 In this way, the decomposition (\ref{decomp1}) can be
written in the Wedderburn's form\cite{Zhelobenko}
\begin{equation}
\sH = \bigoplus_{\mu \in S} \sH_{\mu} \otimes \sM_{\mu}.\label{decomp2}
\end{equation}

Due to Schur lemmas, an operator $O$ in the commutant of the representation $\set{R}(\gG)$ can be
decomposed as follows \cite{MLpovms}
\begin{equation}\label{commutingO}
O=\sum_{\mu} \sum_{i,j=1}^{m_{\mu}}~~ \frac{\Tr[T_{ji}^{(\mu)} O]}{d_{\mu}}~ T_{ij}^{(\mu)},
\end{equation} 
whereas, in terms of the decomposition (\ref{decomp2}) one has
\begin{equation}
O=\oplus_{\mu \in S} \left( I_{\mu} \otimes
O_{\mu}\right),
\end{equation}
$I_{\mu}$ denoting the identity on the representation space
$\sH_{\mu}$, and $O_\mu\in \Bnd{M_{\mu}}$ being a suitable set of operators on the
multiplicity spaces $\sM_{\mu}$. 

In this paper we will consider covariant POVM's with $\Klm= \gG/\gG_0$
where both $\gG$ and $\gG_0$ are compact Lie groups, represented
on the Hilbert space $\sH$ by the unitary representations
$\set{R}(\gG)=\{U_g~|~ g \in \gG\}$ and $\set{R}(\gG_0)=\{U_h~|~h \in
\gG_0\}$. We will denote with $\set{S}$ and $\set{S}_0$ the equivalence classes of irreducible
representations of $\set{R}(\gG)$ and $\set{R}(\gG_0)$ respectively. The constraints 
(\ref{XIposnorm},\ref{XIcomm}) 
can be rewritten in a remarkably simple form using the decompositions
of $\sH$ in irreducible subspaces under the action of $\set{R}(\gG)$ and
$\set{R}(\gG_0)$. In fact, due to the invariance of the Haar measure
$\d g$, the integral in (\ref{XIposnorm}) belongs to the commutant of
$\set{R}(\gG)$. Rewriting the constraint (\ref{XIposnorm}) by using
(\ref{commutingO}), one get easily:
\begin{equation}\label{TRnorm}
\Tr[T_{ij}^{(\mu)} \Xi]=  d_{\mu} ~\delta_{ij},\qquad \forall \mu \in\set{S},\quad \forall
i,j=1,\dots,m_{\mu}. 
\end{equation} 
Moreover, according to (\ref{XIposnorm}) and (\ref{XIcomm}), the operator $\Xi$ must be a positive semidefinite operator in the commutant of $\set{R}(\gG_0)$ (\ref{XIcomm}), then we have
\begin{equation}\label{XIpos-comm}
\Xi =\oplus_{\nu \in S_0}(I_{\nu} \otimes X_{\nu}^{\dag}X_{\nu}),
\end{equation} 
where $X_{\nu}$ is an operator on the multiplicity subspace $\sM_{\nu}$.
\section{Extremal covariant POVM's with a nontrivial stability group}\label{ExtrPovms}
In this section we will classify the extremal points of the convex set $\Conv$ of covariant seeds,
namely the convex set of operators that satisfy both conditions (\ref{XIposnorm}) and (\ref{XIcomm}). 
For the characterization of the extremal points of a convex set we will use the well known method of
perturbations.   We will say that the operator $\Theta$ is a "perturbation" of a given $\Xi\in\Conv$
if and only if there exists an $\epsilon>0$ such that $~\Xi +t \Theta \in \Conv$ for any $t \in
[-\epsilon,\epsilon]$.  With such definition one has that an operator $\Xi$ is extremal if and only
if its unique perturbation is the trivial one, namely if $\Theta$ is a perturbation of $\Xi$ then
$\Theta=0$.

Let's start with a simple lemma which is useful for the characterization of the perturbations of a given seed $\Xi$.
\begin{lemma}\label{PosAndSupp} 
Let $\Xi\in \Bnd{H}$ be a positive semidefinite operator. Then, for any Hermitian $\Theta \in
\Bnd{H}$ the condition 
\begin{equation}\label{epsilon}
\exists \epsilon > 0:  \qquad \forall t \in [-\epsilon,\epsilon] \quad \Xi +t\Theta\geq0
\end{equation}
is equivalent to 
\begin{equation}\label{supports}
\Supp(\Theta) \subseteq \Supp(\Xi).
\end{equation}
\end{lemma}
\Proof Suppose that the condition (\ref{epsilon}) holds. Then for any $|\phi\>\in\Ker(\Xi)$ one
necessarily has $\<\phi|\Theta|\phi\>=0$. Therefore, for any vector $|\psi\>\in\sH$ one
has: 
\begin{equation*}
|\<\psi|\Theta|\phi\>|=\frac{1}{t}|\<\psi|(\Xi+t\Theta)|\phi\>|\leq
\frac{1}{t}\sqrt{\<\psi|(\Xi+t\Theta)|\psi\>~\<\phi|(\Xi+t\Theta)|\phi\>}=0.\end{equation*}
Hence $\Ker{(\Xi)}\subseteq \Ker{(\Theta)}$, implying that
$\Supp{(\Theta)}\subseteq \Supp{(\Xi)}$. Conversely, suppose that
(\ref{supports}) holds. Let's denote by $\lambda$ the smallest nonzero
eigenvalue of $\Xi$ and by $||\Theta||$ the norm of $\Theta$, then
condition (\ref{epsilon}) holds with
$\epsilon=\frac{\lambda}{||\Theta||}$.\qed 
\medskip
Using the previous lemma we can state that an Hermitian operator $\Theta$ is a perturbation for a
given seed $\Xi$ if and only if the following conditions are satisfied:
\begin{eqnarray}
\label{THETAsupp} &\Supp{(\Theta)}\subseteq \Supp{(\Xi)}&\\
\label{THETAnorm} &\Tr[\Theta T_{ij}^{(\mu)}]=0 \qquad&\forall \mu \in S,~ \forall i,j=1,\dots,m_{\mu}\\
\label{THETAcomm} &[\Theta,U_h]=0 \qquad&\forall h\in \gG_0
\end{eqnarray}
(conditions (\ref{THETAnorm}) and (\ref{THETAcomm})  follow directly from the normalization constraints
(\ref{TRnorm}) and (\ref{XIpos-comm})).
\par This set of conditions leads to an interesting property of extremal seeds:
\begin{theorem}
\label{supporti} $\Xi$ is an extremal point of $\Conv$ if and only if for any $\zeta \in \Conv$ one has
\begin{equation}
\Supp(\zeta) \subseteq \Supp(\Xi) ~~ \implies~~ \zeta = \Xi.\label{suppz}
\end{equation}
\end{theorem}
\Proof 
To prove necessity it is sufficient to define $\Theta\doteq \Xi-\zeta$ and note that
it is a perturbation of $\Xi$. In fact, $\Theta$ is in the commutant
of $\set{R}(\gG_0)$, $\Supp(\Theta) \subseteq \Supp(\Xi)$, and
$\Tr[\Theta T_{ij}^{\mu}]=0 \quad \forall \mu \in S, \forall i,j=1,
\dots, m_{\mu}$. But, since $\Xi$ is extremal, then $\Theta$ must be zero. 
\par Viceversa , assume (\ref{suppz}). If $\Theta$ is a perturbation for $\Xi$, then there exists some $t\not =0$ such that $\zeta\doteq \Xi +t \Theta \in \Conv$.  But a perturbation must satisfy (\ref{supports}), then $\Supp(\zeta)\subseteq \Supp(\Xi)$. Using (\ref{suppz}) is then clear that $\Theta=t^{-1}(\zeta-\Xi)=0$.\qed
\medskip
The proposition tells us that extremal seeds have "minimal support", in the sense that there is no element $\zeta\in \Conv$ with $\Supp{(\zeta)}\subseteq \Supp(\Xi)$ which is different from $\Xi$. 
\medskip
\begin{theorem}
\label{XIpert}   Let be $\Xi \in \Conv$.
Write $\Xi$ in the form (\ref{XIpos-comm}).  Then an operator $\Theta$
is a perturbation of $\Xi$ if and only if
\begin{equation}\label{pert1}
\Tr[\Theta T_{ij}^{(\mu)}]=0 \qquad \forall \mu \in S,~\forall i,j =1, \dots, m_{\mu}
\end{equation}
and $\Theta$ can be written as follows
\begin{equation}\label{pert2}
\Theta = \oplus_{\nu \in S_0} \left(I_{\nu} \otimes X_{\nu}^{\dag}A_{\nu}X_{\nu}\right),
\end{equation}
with $X_\nu\in\Bnd{M_\nu}$ and $A_{\nu}\in \Bndd{\Rng(X_{\nu})}$ Hermitian
$\forall\nu\in\set{S}_0$. 
\end{theorem}
\Proof  Suppose $\Theta$ is a perturbation.  Condition (\ref{THETAnorm}) is the same as
(\ref{pert1}). Due to condition 
(\ref{THETAcomm}), $\Theta$ must be an Hermitian operator in the commutant
of $\set{R}(\gG_0)$, then we can write it in the block form $\Theta=
\oplus_{\nu \in\set{S}_0} (I_{\nu} \otimes O_{\nu})$, with each $O_{\nu}\in\Bnd{M_{\nu}}$
Hermitian. Moreover, condition (\ref{THETAsupp}) along with (\ref{XIpos-comm}) imply that
each operator $O_{\nu}$ must have $\Supp(O_{\nu}) \subseteq
\Supp(X_{\nu}^{\dag}X_{\nu})= \Supp(X_{\nu})$. Using the singular value decomposition
$X_{\nu}=\sum_{i=1}^{r_{\nu}} \lambda_{i}^{(\nu)}
|w_i^{(\nu)}\>\<v_i^{\nu}|$ ($\{|v_i^{\nu}\>\}$ and
$\{|w_i^{(\nu)}\>$ are orthonormal bases for $\Supp(X_{\nu})$ and
$\Rng(X_{\nu})$ respectively) one can see that any Hermitian operator $O_{\nu}$ with
$\Supp(O_{\nu})\subseteq\Supp(X_{\nu})$ admit the decomposition
$O_{\nu}=X_{\nu}^{\dag}A_{\nu}X_{\nu}$, with $A_{\nu}$ Hermitian operator in 
$\Bndd{\Rng(X_{\nu})}$. Conversely, if both conditions (\ref{pert1}) and (\ref{pert2}) hold, then 
conditions (\ref{THETAsupp}--\ref{THETAcomm}) are obviously fulfilled. \qed
\begin{theorem}\label{th:iff}
Let be $P_{\nu}$ the projection operator onto the subspace $\sH_{\nu} \otimes \sM_{\nu}\subseteq
\sH$ corresponding to the class $\nu \in S_0$. An operator $\Xi \in \Conv$ written in the form 
$\Xi =\oplus_{\nu \in S_0}(I_{\nu} \otimes X_{\nu}^{\dag}X_{\nu})$ 
is extremal if and only if
\begin{equation}\label{iff}  \oplus_{\nu \in S_0}\Bndd{\Rng(X_{\nu})}= \Span \{F_{ij}^{(\mu)}
~|~\mu\in S,~ i,j=1, \dots, m_{\mu}\}, \end{equation}
where 
\begin{equation*} F_{ij}^{(\mu)} \doteq \oplus_{\nu \in S_0}~ X_{\nu}~\Tr_{\sH_{\nu}}\left[P_{\nu} T_{ij}^{(\mu)} P_{\nu}\right]~X_{\nu}^{\dag}.
\end{equation*}  
\end{theorem}
\Proof  Using the characterization of Theorem \ref{XIpert}, we know that
  $\Xi$ is extremal if and only if for any operator $\Theta$
  satisfying $(\ref{pert1})$ and $(\ref{pert2})$ one has
  $\Theta=0$. Let's take $\Theta$ in the form (\ref{pert2}), and
  rewrite the direct sum as an ordinary sum 
\begin{equation}\Theta=\sum_{\nu \in
  S_0} P_{\nu} \left( I_{\nu} \otimes
  X_{\nu}^{\dag}A_{\nu}X_{\nu}\right) P_{\nu},
\end{equation} using the projectors  $P_{\nu}$ onto $\sH_{\nu} \otimes M_{\nu}$. Using invariance
of trace under cyclic permutations, we can write
\begin{equation}
\begin{split}
 \Tr\left[ \Theta~T^{(\mu)}_{ij}\right] =& \sum_{\nu \in S_0} \Tr\left[ (I_{\nu} \otimes A_{\nu}) (I_{\nu}
 \otimes X_{\nu}) P_{\nu} T^{(\mu)}_{ij} P_{\nu}(I_{\nu} \otimes
 X_{\nu}^{\dag})\right]\\=&\sum_{\nu \in S_0}\Tr \left[
 A_{\nu}~~ X_{\nu} \Tr_{\sH_{\nu}}
 [P_{\nu} T^{(\mu)}_{ij} P_{\nu}]X^{\dag}_{\nu}\right].
\end{split}
\end{equation}
Define the space $\sR\doteq \oplus_{\nu \in S_0}\Rng(X_{\nu})$ and  denote as
$\oplus_{\nu \in S_0} \Bndd{\Rng(X_{\nu})}$ the linear space of
operators acting on $\sR$ which are block diagonal on the subspaces $\Rng(X_{\nu})$, 
$\nu\in\set{S}_0$. Then, the extremality condition for $\Xi$
becomes: for any Hermitian operator $A \in \oplus_{\nu \in S_0} \Bndd{\Rng(X_{\nu})}$ one has
\begin{equation}
\Tr\left[ A F^{(\mu)}_{ij}\right]=0 \quad \forall \mu \in\set{S},~\forall i,j=1, \dots,m_{\mu} \quad \implies \quad A=0.
\end{equation}
In terms of the Hilbert-Schmidt product $(A,B)\doteq \Tr[A^{\dag}B]$ this condition says that the
unique Hermitian operator $A\in
\oplus_{\nu \in S_0} \Bndd{\Rng(X_{\nu})}$ which is orthogonal to the
whole set of operators $\set{F}\doteq\{F^{(\mu)}_{ij}~~|~~ \mu \in\set{S},~
i,j=1, \dots, m_{\mu}\}$ is the null operator. Orthogonality to
the set $\set{F}$ is equivalent to orthogonality to the set of
Hermitian operators
$\set{F}'=\{(F^{(\mu)}_{ij}+F_{ji}^{(\mu)})~,~i(F_{ij}^{(\mu)}-~F_{ji}^{(\mu)})~|~\mu
\in\set{S},~i,j=1,\dots m_{\mu}\}$. Such orthogonality holds if and only if
$\set{F}'$ is a spanning set for the real space of Hermitian
operators in $\oplus_{\nu \in S_0}\Bndd{\Rng(X_{\nu}}$. Nevertheless, using the
Cartesian decomposition we see that any complex block operator 
$O\in\oplus_{\nu \in S_0} \Bndd{\Rng(X_{\nu})}$ can be written as sum of
two Hermitian ones, whence the extremality condition is equivalent to
$\Span(\set{F}')=\oplus_{\nu \in S_0}
\Bndd{\Rng(X_{\nu}}$. Finally, the observation
$\Span(\set{F}')=\Span(\set{F})$ completes the proof.
\qed
Notice that for trivial stability group $\gG_0=\{e\}$ ($e$ denotes the identity element), we recover
the characterization of \cite{extPOVMandQO}: there, one has indeed a single equivalence class
$\bar\nu$ in $\set{S}_0$ with one-dimensional representation space $\sH_{\bar\nu}$, so that the
whole Hilbert space $\sH$ is isomorphic to the multiplicity space $\sM_{\bar\nu}$ and the
extremality condition (\ref{iff}) reduces to $\Span\{X T^{(\mu)}_{ij}X^{\dag}~|~ \mu 
\in\set{S},~i,j=1, \dots, m_{\mu}\}= \Bndd{\Rng(X)}$. 
\begin{corollary}\label{rankone-extr} Any rank-one seed is extremal.\end{corollary}
\Proof  Let be $\Xi$ a rank-one seed. In this case there is only one class $\nu_0$ in the
decomposition (\ref{XIpos-comm}) of $\Xi$ (otherwise $\Xi$ could not
have unit rank), and the space $\Bndd{\Rng(X_{\nu_0})}$ to be spanned
is one dimensional, whence the condition (\ref{iff}) is always
satisfied. \qed An alternative proof of Corollary \ref{rankone-extr}
follows by observing that any rank-one element of the cone $\set{D}$ of positive
semidefinite operators is necessarily extremal for such cone: since the convex set
$\Conv$ is a subset of $\set{D}$, a rank-one seed $\Xi\in\Conv$ is necessarily an extreme point of 
$\Conv$.
\begin{corollary}\label{RankBound}
Let $\Xi\in \Conv$ be an extremal seed and write it in the form $\Xi =\oplus_{\nu \in S_0}(I_{\nu}
\otimes X_{\nu}^{\dag}X_{\nu})$. Define $r_{\nu}\doteq \rnk(X_{\nu})$. Then 
\begin{equation}\label{ranks}
\sum_{\nu \in\set{S}_0} r_{\nu}^2 \leq \sum_{\mu \in\set{S}}  m_{\mu}^2.
\end{equation} 
\end{corollary}
\Proof This relation follows directly from the extremality condition by noting
 that the left hand side is the dimension of the complex linear space
 of block operators $\oplus_{\nu \in\set{S}_0} \Bndd{\Rng(X_{\nu})}$, while
 the right hand side is the cardinality of the spanning set
 $\set{F}=\{F_{ij}^{(\mu)}~|~\mu \in\set{S},~i,j=1, \dots, m_{\mu}\}$.\qed
In Section \ref{s:examples} we will see an explicit example of extremal POVM which achieves this bound.
\section{Extremal  POVM's and optimization problems}\label{OptExtr}
A crucial step in a quantum estimation approach is the optimization of the estimation strategy for a
given figure of merit. This consists in finding the POVM which maximizes some linear (more generally
concave) functional $\mathcal{F}$---e. g. the average fidelity of the estimated
state with the true one. Then, the convex structure of the set of POVM's
plays a fundamental role in this problem, since, due to concavity of $\mathcal{F}$, one can restrict
the optimization procedure to the extremal POVM's only.

In the covariant case, the problem resorts to optimize the state estimation in the orbit $\{U_g\rho
U_g^{\dag}~|~ g\in \gG\}\simeq\gG/\gG_0$ 
of a given state $\rho$ under the action of a group
$\gG$, $\gG_0$ being the stability group of $\rho$. The optimization typically is
the maximization of a linear functional corresponding to the average value of a positive function
$f(x,x_*)$, where the average is taken over all the couples $(x,x_*)$ of measured and true
values $x,x_*\in\Klm\doteq\gG/\gG_0$, respectively. The joint probability density 
$p(x,x_*)$ is connected to the conditional density $p(x|x_*)$ given
by the Born rule via Bayes, assuming an \emph{a priori} probability distribution
of the true value $x_*$. In the covariant problem the function $f$ enjoys the invariance
property $f(gx,gx_*)=f(x,x_*)$ $\forall g \in \gG$, and is taken as a decreasing function of the
distance $|x-x_*|$ of the measured value $x$ from the true one $x_*$. In the case of compact $\gG$
one can assume a uniform \emph{a priori} distribution for $x_*$ values, so that the functional
corresponding to the average can be written as follows 
\begin{eqnarray}
\label{average}\mathcal{F}_{\rho}[\Xi] &=& \int_{\gG} \d g \int_{\gG} \d g_*~~ f(gx_0,g_*x_0)~ \Tr[U_{g_*} \rho U_{g_*}^{\dag} U_g \Xi U_g^{\dag}]\\
&=& \int_{\gG}\d g ~~ f(x_0,gx_0)~ \Tr[U_g \rho U_g^{\dag} \Xi],
\end{eqnarray}
where $x_0$ is the equivalence class containing the identity. In the
following, we will consider as the prototype optimization problem
the maximization of the likelihood functional\cite{helstrom,holevo}
\begin{equation}
\mathcal{L}_{\rho}[\Xi]\doteq\Tr[\rho \Xi],
\end{equation}
corresponding to the choice $f(x,x_*)=\delta(x-x_*)$ in Eq.(\ref{average}). Maximizing 
$\mathcal{L}_{\rho}[\Xi]$ means maximizing the probability density that the measured value $x$
coincides with the true value $x_*$. For such estimation strategy the optimization problem
has a remarkably simple form, enabling a general treatment for a large class of group
representations \cite{MLpovms}. Moreover, the solution of the maximum likelihood is formally
equivalent to the solution of any optimization problem with a positive (which, a part from an
additive constant, means bounded from below) summable function $f(x,x_*)$. Indeed, we can define the
map 
\begin{equation}
\map{M}(\rho) = k^{-1} \int_{\gG}\d g~~ f(x_0,gx_0)~ U_g \rho U_g^{\dag},\label{maplike}
\end{equation}
where $k= \int_{\gG} \d g~~f(x_0,gx_0)$. This map is completely positive, unital and trace preserving, and, in particular, $\map{M}[\rho]$ is a state.  With this definition, we have 
\begin{equation}\mathcal{F}_{\rho}[\Xi] =  k~  \mathcal{L}_{\map{M}(\rho)}[\Xi],
\end{equation} 
whence the maximization of $\mathcal{F}_{\rho}$ is equivalent to the maximization of the likelihood for the transformed state $\map{M}(\rho)$. 
\par Essentially all optimal covariant measurements known in the literature are represented by
rank-one operators. The rank-one assumption often provides a useful instrument for simplifying
calculations.  Nevertheless, as we will show in the following,  the occurrence of 
POVM's with rank grater than one is unavoidable in some relevant situations.
\begin{proposition}
For any $\Xi \in \Conv$,
\begin{equation}
\rnk[\Xi] \geq
\max_{\mu \in\set{S}}\left(\frac{m_{\mu}}{d_{\mu}}\right).
\end{equation} 
\end{proposition}
\Proof  Let's decompose $\sH$ into irreducible subspaces for the representation $\set{R}(\gG)$ of
$\gG$ as follows 
 \begin{equation}\sH=\oplus_{\mu \in S}\oplus_{i=1}^{m_{\mu}}
\sH_i^{(\mu)}.
\end{equation} 
Take an orthonormal basis
$\set{B}^{(\mu)}_i=\{|(\mu,i), n\>~|~n=1, \dots, d_{\mu}\}$ for each
subspace $\sH_i^{(\mu)}$ in such a way that $|(\mu,i), n\>= T^{(\mu)}_{ij}|(\mu,j), n\>$
for any $n$, $T^{(\mu)}_{ij}: \sH_j \to \sH_i$ being the invariant isomorphism which intertwines the
equivalent representations $(\mu,i)$ and $(\mu,j)$.  Diagonalize $\Xi$ as 
\begin{equation}
\Xi =\sum_{k=1}^{\rnk(\Xi)} |\eta_k\>\<\eta_k|
\end{equation}
and write 
\begin{equation}
|\eta_k\>=\sum_{\mu \in\set{S}}\sum_{i=1}^{m_{\mu}} \sum_{n=1}^{d_{\mu}} c^k_{(\mu,i),n}~~ |(\mu,i),n\>.
\end{equation}
Since $\<\eta_k| T^{(\mu)}_{ij}|\eta_k\>=\sum_{n=1}^{d_{\mu}} c^{k*}_{(\mu,i),n} c^{k}_{(\mu,j),n}$, the normalization constraints (\ref{TRnorm}) become
\begin{equation}
\sum_{k=1}^{\rnk(\Xi)} \sum_{n=1}^{d_{\mu}}~  c^{k*}_{(\mu,i),n}c^{k}_{(\mu,j),n} = d_{\mu}~ \delta_{ij}.
\end{equation}
This relation implies that for any $\mu \in\set{S}$ the vectors
$\{\vec{c}_{(\mu,i)}~|~i=1, \dots,m_{\mu}\}$ defined by
$ (\vec{c}_{(\mu,i)})_{k,n} \doteq c^k_{(\mu,i),n}$ are orthogonal:
 since they are $m_{\mu}$ orthogonal vectors in a linear space whose
dimension is $d_{\mu}\times\rnk(\Xi)$, it follows that $m_{\mu} \leq d_{\mu} \times\rnk(\Xi)$, 
hence $\rnk(\Xi)\geq \frac{m_{\mu}}{d_{\mu}}\quad \forall \mu \in\set{S}$. \qed 
\medskip
Summarizing, every times $m_{\mu}>d_{\mu}$ for some class $\mu \in\set{S}$, a covariant POVM 
cannot be represented by a rank-one seed.  
\medskip
\par The previous proposition exhibits a structural reason for which, in
the presence of equivalent representations, the set $\Conv$ of
covariant seeds may contain only elements with rank greater then one. On
the other hand, in the following we will discuss the occurrence of
covariant POVM's with rank greater than one in explicit optimization
problems, independently of the presence of equivalent representations.
\begin{proposition}
\label{uniqueOptPOVM} Let be $\Xi$ an extremal point of $\Conv$. Denote by $P$ the projector onto $\Supp(\Xi)$, and let $r\doteq\rnk(P)$. Then $\Xi$ is the unique seed which maximizes the likelihood for the state $\rho=\frac{P}{r}$.
\end{proposition}
\Proof 

First, we need to prove that $\Xi$ commutes with the representation
$\set{R}(\gH_0)\doteq \{U_k ~|~ k\in\gH_0\}$, where $\gH_0$ is the
stability group of $\rho$, defined by  $[\rho,U_k]=0 \quad \forall k \in \gH_0$. Define the group average
\begin{equation}
\xi \doteq \frac{\int_{\gH_0}\d h\, U_h\Xi U_h^\dag}{\int_{\gH_0}\d h}~.
\end{equation}
Since $\set{R}(\gH_0)$ is the stability group of the projector onto $\Supp(\Xi)$, clearly $\Supp(\Xi)$ is invariant under $\set{R}(\gH_0)$, whence $\xi$ satisfies
$\Supp(\xi) \subseteq
\Supp(\Xi)$. Moreover, using the invariance of the Haar measure it is easy to see that $\xi$ commutes with $\set{R}(\gH_0)$.  Finally, $\xi$ is an element of $\Conv$. In fact, it is positive semidefinite, satisfies $(\ref{TRnorm})$ and commutes with $\set{R}(\gG_0)$---the stability group of $\Xi$---which is by definition a subset of $\set{R}(\gH_0)$. Since $\Xi$ is extremal, using Theorem \ref{supporti} we can conclude that $\Xi=\xi$, whence $\Xi$ commutes with $\set{R}(\gH_0)$.
\par Let's prove now optimality. For any arbitrary seed $\zeta\in \Conv$, the following bound holds:
\begin{equation}
\mathcal{L}_{\rho}[\zeta]= \Tr[\rho \zeta]= \frac{\Tr[P \zeta]}{r} \leq \frac{\Tr[\zeta]}{r}= \frac{\dim(\sH)}{r},
\end{equation}
where the last equality follows from the normalization constraints
(\ref{TRnorm}). Clearly $\Xi$ achieves the bound, whence it is
optimal. Notice that the inequality $\Tr[P
\zeta] \leq \Tr[\zeta]$ becomes equality if and only of $\Supp(\zeta)
\subseteq \Supp(\Xi)$, then using Theorem \ref{supporti} we can see
that $\Xi$ represents the unique optimal POVM. \qed 
\medskip
Consider now a density matrix $\sigma$ with support in the orthogonal complement of $\Supp(\Xi)$, 
and consider the randomization 
\begin{equation}\label{randomized}
\rho=(1-\alpha) \frac{P}{r} + \alpha
\sigma,
\end{equation}
with $0\leq \alpha \leq 1$. In the following we prove that, for sufficiently small $\alpha>0$, $\Xi$
is still optimal for the maximum likelihood strategy. In other words, the extremal POVM represented
by $\Xi$ is stable under randomization, and the same measuring apparatus can be used for a larger
class of mixed states.  
\begin{proposition}
\label{stabilityPOVM}  Consider the randomized state $\rho$ in (\ref{randomized}) and denote by $\bar{q}$ the maximum eigenvalue of $\sigma$. If $\alpha < \frac{1}{1+ r \bar{q}}$, then $\Xi$ is the unique seed which maximizes the likelihood for the state $\rho$.
\end{proposition}
\Proof First, notice that $\Xi$ commutes with the representation $\set{R}(\gH_0)$ of the stability
group of $\rho$. This follows from the observation that the condition $\alpha <
\frac{1}{1+r\bar{q}}$ implies that $\frac{1-\alpha}{r}$ is strictly the
largest eigenvalue of $\rho$. Then, $P$ is the projector on the
eigenspace with maximum eigenvalue of $\rho$, while,  for any $h\in \gG$, $P_h \doteq U_h P
U_h^{\dag}$ is the projector on the eigenspace with maximum eigenvalue
of $\rho_h \doteq U_h \rho U_h^{\dag}$. If $h \in \gH_0$ then it must
be $\rho_h =\rho$, and, necessarily, $P_h=P$. Therefore $\gH_0$ is a
subgroup of the stability group of $P$. But $\Xi$ commutes with the
representation of the stability group of $P$, as proven in Proposition
\ref{uniqueOptPOVM}, then it commutes also with $\set{R}(\gH_0)$.
\par Now we prove optimality of $\Xi$.  Let's
denote by $Q$ the projection onto $\Supp(\sigma)$. The following bound
holds
for any $\zeta \in
\Conv$:
\begin{eqnarray}
\mathcal{L}_{\rho}[\zeta]&=& \frac{(1-\alpha)}{r} \Tr[P\zeta] + \alpha \Tr[\sigma \zeta]\\
&\leq& \frac{(1-\alpha)}{r} \Tr[P\zeta]+ \alpha\bar{q} \Tr[Q
\zeta]\\ \label{ineq1} &\leq& \frac{(1-\alpha)}{r} \Tr[(P+Q)\zeta]\\
\label{ineq2}&\leq& \frac{(1-\alpha)}{r} \Tr[\zeta]
= \frac{(1-\alpha)}{r} \dim(\sH).
\end{eqnarray} 
This bound is achieved by $\Xi$, proving its optimality. Notice that  $\Xi$ is the unique optimal
seed. In fact, equality in (\ref{ineq1}) is attained if and only if $\Tr[Q\zeta]=0$, namely when $\Supp(Q)\subseteq
\Ker(\zeta)$, while in (\ref{ineq2}) equality is 
attained if and only if 
$\Supp(\zeta)\subseteq \Supp(P)\oplus \Supp(Q)$. Therefore the bound is
achieved if and only if $\Supp(\zeta) \subseteq \Supp(P)=\Supp(\Xi)$,
implying $\zeta=\Xi$.\qed 
\section{Examples}\label{s:examples}\subsection{Extremal POVM's with a non trivial stability group}
\subsubsection{} Consider the group of rotations, represented in a  
$(2j+1)$-dimensional Hilbert space $\sH_j$ by the irreducible representation $R_{{\bf n},
\varphi}\doteq e^{i\varphi {\bf n}\cdot{\bf j}}$, where $\varphi$ is an angle, ${\bf n}$ is a unit-vector, and ${\bf j}
\doteq (j_x,j_y,j_z)$ is the angular momentum operator. In this case
a covariant estimation in the orbit of a pure state $|\psi\>$ generally 
may involve a nontrivial stability group. This is actually the case when
$|\psi\> \doteq |j m\>_{\bf n_0}$, is an eigenvector
of ${\bf n_0 \cdot j}$ for some unit vector ${\bf n_0}$. Clearly in
such case the stability group $\gG_0$ consists of rotations around ${\bf n_0}$, and the state
estimation in the orbit reduces to the estimation of a rotated direction 
${\bf n'}$.  The same situation arises for any state $\rho$ mixture of eigenvectors of ${\bf n_0
  \cdot j}$. Without loss of generality, let's take ${\bf n_0}$ as the direction of the $z$-axis, and
write $\rho=\sum_{m=-j}^{j} p_m |jm\>\<jm|$ with $p_m \geq 0 \quad \forall
m$. Let's denote by $P$ the projector onto $\Supp(\rho)$, and take
$\bar{m}$ such that $p_{\bar{m}}=\max_{m}\{p_m\}$.  Then, since
\begin{equation*}
\Tr[\rho  \zeta] \leq p_{\bar{m}} \Tr[P \Xi] \leq p_{\bar{m}}\Tr[\Xi] =p_{\bar{m}} (2j+1),
\end{equation*} 
one has that $\Xi = (2j+1) |j\bar{m}\>\<j\bar{m}|$ is the
optimal POVM. Notice that such POVM commutes with the stability group
$\set{R}(\gG_0)$ and is extremal, as a consequence of Corollary \ref{rankone-extr}.  
\subsubsection{}
Consider the group $\mathbb{SU}(d)$ of unitary $d
 \times d$ matrices with unit determinant, acting on the space $\sH \doteq \Cmplx^d$. It is easy to
 see that each vector $|\psi\> \in \sH$ has a nontrivial stability group $\gG_0\equiv
\mathbb{U}(d-1)$. In fact, by 
 introducing an orthonormal basis $\set{B}_{\perp} \doteq
\{|n\>~|~n=1, \dots ,d-1\}$ for the orthogonal complement $\sH^{\perp}$ of the line $\Span\{|\psi\>\}$,
and the basis $\set{B} \doteq |\psi\> \cup \set{B}_{\perp}$
for $\sH$, the stability group $\gG_0$ consists on matrices of the form
\begin{equation}\label{block}
U_{h} = \left( \begin{array}{l|lll}
                        \omega_h &&{\bf 0}\\
\hline
                        {\bf 0} && V_h\\
                \end{array}~
        \right),
\end{equation}
where $\omega_h\in \Cmplx, \ |\omega_h|=1$, and $V_h$ is a unitary $(d-1)\times (d-1)$
matrix with $\Det(V_h)=\omega_h^*$.   
Let's consider now the tensor
representation $\set{R}(\gG)=\{U_g^{\otimes 2}~|~ U_g \in\mathbb{SU}(d) \}$ on the space
$\sH^{\otimes 2}$. This representation has two irreducible subspaces, the symmetric and the
antisymmetric ones $\sH_+$ and $\sH_-$, with dimensions $d_+=\frac{d(d+1)}{2}$ and $d_-=\frac{d(d-1)}{2}$ respectively. Denote by $P_+$ and $P_-$ the projectors on $\sH_+$ and
$\sH_-$. Let's apply the representation $\set{R}(\gG)$ on the state $|\psi\>^{\otimes
  2} \in \sH^{\otimes 2}$. Clearly the stability group is the 
same $\gG_0$ as before, and it is represented by 
$\set{R}(\gG_0)=\{U_h^{\otimes 2} ~|~h\in\gG_0\}$. It is easy
to see that $\set{R}(\gG_0)$ contains five irreducible
components, carried by the subspaces $\sH_1=
\Span\{|\psi\>^{\otimes 2}\}$, $\sH_2= \Span\{|\psi\>\} \otimes
\sH^{\perp}$ , $\sH_3 = \sH^{\perp} \otimes \Span\{|\psi\>\}$,
$\sH_4= P_+(\sH^{\perp~\otimes 2})$, and $\sH_5=P_-(\sH^{\perp~
\otimes 2})$. Notice that $\sH_2$ and $\sH_3$ carry equivalent
representations, corresponding to a two dimensional multiplicity
space. An example of extremal POVM is given by
\begin{equation*}
\Xi = \frac{d(d+1)}{2}~
|\psi\>\<\psi|^{\otimes 2} \oplus \frac{d}{d-2}~  P_- Q P_-,
\end{equation*}
where $Q$ is the projection on $\sH^{\perp~ \otimes 2}$. Since the two
summands are proportional to $|\psi\>\<\psi|^{\otimes 2}$ and $P_- Q P_-$, which are the projectors on $\sH_1$ and
$\sH_5$ respectively, then $\Xi$ belongs to the commutant of $\set{R}(\gG_0)=\{U_h^{\otimes
2}~|~h\in\gG_0\}$.  Notice that the subspaces $\sH_1$ and $\sH_5$ have multiplicities 
$m_1=m_5=1$, corresponding to one-dimensional multiplicity spaces 
$\sM_1 \equiv \sM_5 \equiv \Cmplx$ (whence the partial traces over $\sH_{1,5}$ will be
$c$-numbers). Moreover, using the fact that $\Tr_{\sH_1}[P_+]=1$,
$\Tr_{\sH_1}[P_-]=0$, $\Tr_{\sH_5}[P_+]=0$,
$\Tr_{\sH_5}[P_-]=\frac{(d-1)(d-2)}{2}$ one can check extremality using
the condition (\ref{iff}). Let's observe that in this example we have $r_1=r_5=1$ and $m_+=m_-=1$,
where $r_1$ and $r_5$ are defined as in Corollary \ref{RankBound}, while $m_+$ and $m_-$ are the
multiplicities of the two irreducible representations of $\set{R}(\gG)$. Then the bound of
(\ref{ranks}) is saturated.  Finally, we remark that this POVM is 
optimal for discriminating states in the orbit of $|\psi\>^{\otimes
2}$ \cite{MLpovms}, in the orbit of $\rho= \frac{1}{r}
\left(|\psi\>\<\psi|^{\otimes 2} + P_- Q P_- \right)$ where $r = 1+
\frac{(d-1)(d-2)}{2}$ because of Proposition \ref{uniqueOptPOVM}, and also in the orbit of any
randomization $\rho'=(1-\alpha)\rho + \alpha \sigma$ where $\sigma$ is density matrix with
$\Supp(\sigma)\subseteq \Ker(P)$, and $\alpha < \frac{1}{1+r}$, 
because of  Proposition \ref{stabilityPOVM}. 
\subsection{Extremal POVM's with rank greater than one}
\subsubsection{} Consider the Abelian group $\gG=\mathbb{U}(1)$ of phase shifts, acting in the space
$\sH= \Cmplx^{d}$ by the representation $\set{R}(\gG)=\{U(\varphi) = \exp(i\varphi
N\}~|~ \varphi \in [-\pi, \pi]\}$, where the generator $N$ is given by $N= \sum_{n=0}^{d-1}
n~|n\>\<n|$ for some orthonormal basis $\{|n\>~|~n=0,1, \dots, d-1\}$.
The stability  group $\gG_0$ may be either the whole $\mathbb{U}(1)$ (for $\rho$ diagonal on the 
eigenstates of the generator), or a discrete subgroup $\gG_0 =
\mathbb{Z}_k$ for some integer $k$, including the case $k=1$ of
trivial stability group. We exclude the degenerate case $\gG_0=\mathbb{U}(1)$ of shift invariant
states. The parameter space $\Klm=\mathbb{U}(1)/\mathbb{Z}_k$ will be a circle, parametrized by an
angle $\theta \in [-\pi,\pi]$, and the action of a group element
$g(\varphi)\in \gG$ on an element $\theta\in\Klm$ 
will be given by $g(\varphi)~\theta= \theta + k \varphi$.
\par Due to constraint (\ref{TRnorm}), a seed $\Xi$ is represented in the eigenbasis of the generator by a correlation matrix, namely by a positive semidefinite matrix with unit diagonal entries. Viceversa, any correlation matrix corresponds to a seed in the case of trivial stability group $\gG_0$. In \cite{LiTam} one can find a constructive method which provides extremal correlation matrices with rank $r>1$: here we show that any of such matrices can be viewed as the optimal seed for the estimation problem in the orbit of
a particular state. Let us choose as optimality criterion the
maximization of the average value of a positive summable function
$f: \Klm \times \Klm \to \mathbb{R}_+$ depending only on the difference
$\theta-\theta_*$ between the measured and the true value.  Suppose $\rho$ a state with stability group $\gG_0=\mathbb{Z}_k$. As we noted at the beginning of section \ref{OptExtr}, the
maximization of 
$\set{F}_{\rho}[\Xi]$---the average value of $f(\theta-\theta_*)$---corresponds to the maximization of the
likelihood $\mathcal{L}_{\map{M}(\rho)}[\Xi]$ for the transformed state
$\map{M}(\rho)= f_0^{-1}\int_{-\pi}^{\pi}\frac{\d \varphi}{2\pi} f(-k\varphi)
U_{\varphi}\rho U_{\varphi}^{\dag}$ (from Eq. (\ref{maplike})). Notice that the map $\map{M}$ 
is trivially covariant---i.e. $\map{M}(U_\phi\rho U_\phi^\dag)=U_\phi\map{M}(\rho)
U_\phi^\dag$----since the group is abelian. For simplicity here we require that the map $\map{M}$ is
invertible, whence also $\map{M}^{-1}$ is covariant and trace-preserving (but generally not
positive). Covariance of $\map{M}$ implies that the stability group of $\map{M}(\rho)$
contains the stability group of $\rho$, and covariance of $\map{M}^{-1}$ implies the reverse
inclusion, whence the stability group is not changed by the maps.
\par Let's take now an extremal correlation matrix $\Xi$ 
with $\rnk(\Xi)= r\geq 1$ and denote by $P$ the projector onto
$\Rng(\Xi)$. Using Proposition \ref{uniqueOptPOVM}, we can see  that $\Xi$ commutes with the representation $\set{R}(\gH_0)$, where $\gH_0$ is the stability group of $P$.   Call $\lambda$ the modulus of the minimum eigenvalue of
$\mathcal{M}^{-1}(\frac{P}{r})$, then 
\begin{equation*}
\rho= \frac{\lambda}{1+d\lambda }~
I+ \frac{1}{1+d\lambda} \mathcal{M}^{-1}(\frac{P}{r})
\end{equation*} is a density operator. Notice that the stability group $\gG_0$ of $\rho$ is the same stability group of $\mathcal{M}^{-1}(P)$, which coincides with $\gH_0$, the stability group of $P$. Therefore $\Xi$ commutes with the representation $\set{R}(\gG_0)$. It is
easy to show that $\Xi$ is the unique seed commuting with $\set{R}(\gG_0)$ which is also optimal for the estimation
of states in the orbit of $\rho$. In fact, for any $\zeta$ in the convex set  $\Conv$ of the seeds with stability group $\gG_0$, we have
\begin{eqnarray*}
\set{F}_{\rho}[\zeta]&=& f_0 \Tr[\zeta\mathcal{M}(\rho)]=f_0\left( \frac{\lambda}{1+d\lambda} \Tr[\zeta] + \frac{1}{r(1+d\lambda)} \Tr[\zeta P]\right)\\
&\leq& f_0 \left(\frac{d}{r}\right)~\left(\frac{1+r\lambda}{1+d\lambda}\right)
\end{eqnarray*}
This bound is achieved choosing $\zeta=\Xi$, 
moreover, as in Proposition \ref{uniqueOptPOVM}, we 
can observe that  the functional $\Tr[\zeta P]$ with $\zeta \in\Conv$ is maximum if and only if $\zeta=\Xi$, then
the maximum is unique.  
\subsubsection{} We provide now an example with a
non-compact group represented in an infinite dimensional Hilbert
space. This example is out of the general treatment of the present paper---which considers only
finite dimensions---and is given only with the purpose of showing that our results could be
generalized to infinite dimensions, however at the price of much more technical proofs.  

Take $\sH$ as the Fock space, and consider the projective representationon $\sH$ of the 
group of translations on the complex plane $\Cmplx$ in terms of the Weyl-Heisenberg operators
$\set{R}(\gG)=\{D(\alpha)=e^{\alpha a^{\dag}-\bar{\alpha}a}~|~ \alpha \in \Cmplx\}$, where $[a,a^{\dag}]=1$.  Here we will
 consider the 2-fold tensor representation $\{D(\alpha)^{\otimes 2}~|~
 \alpha \in \Cmplx\}$ on $\sH^{\otimes 2}$.  Using the unitary
 operator $V=e^{\frac{\pi}{4}(a_1a_2^{\dag}-a_1^{\dag}a_2)}$, one can
 write $D(\alpha)^{\otimes 2}= V (D(\sqrt{2}\alpha) \otimes I)V^{\dag}$ and
 see that the irreducible subspaces of this representation are $\sH_n =
 V (\sH
\otimes \Span(|\phi_n\>)$, $\{|\phi_n\>~|~ n=1,2, \dots \infty \}$ any 
orthonormal basis for $\sH$. All these subspaces carry equivalent
representations, the isomorphism between $\sH_m$ and $\sH_n$ being
\begin{equation}
T_{mn} =V(I \otimes |\phi_m\>\<\phi_n|)V^{\dag}.\label{Tmn}
\end{equation}
In terms of these isomorphisms, the normalization constraints
(\ref{TRnorm}) for a seed operator become \cite{MLpovms}
\begin{equation}
\Tr[T_{mn} \zeta] = 2 \delta_{mn}
\end{equation}  
Notice that the number 2 in this formula has nothing to do with the
dimension of $\sH_n$ which is infinite: in the non-compact case the
dimensions are replaced by positive numbers depending only on the equivalence class of
representations. In principle, since the space $\sH^{\otimes 2}$ is
infinite dimensional, there is the possibility of extremal covariant
POVM's with an infinite rank. Actually we can provide the remarkable
example
\begin{equation}
\Xi= 2~V (|0\>\<0|\otimes I)V^{\dag},
\end{equation}
where $|0\>$ is the vacuum state of the Fock basis $\{|m\>~|~ a^{\dag}a|m\>=m|m\>\}$. The corresponding POVM can be realized by averaging the outcomes of two independent measurement with 
$\Xi_1= |0\>\<0|\otimes I$ and $\Xi_2= I \otimes |0\>\<0|$  \cite{MLpovms}, which in 
quantum optics correspond to two heterodyne measurements \cite{bilkent}.  

We can observe that $\Xi$ is maximizes the likelihood functional for any state of the form $\rho= V
(|0\>\<0| \otimes \sigma) V^{\dagger}$, where 
$\sigma= \sum_{n=0}^{\infty} p_n |\phi_n\>\<\phi_n|$, is a mixed state
with $p_n >0~\forall n$. In fact, for any seed $\zeta$, one has the bound
\begin{equation}
\begin{split}
\Tr[V(|0\>\<0|\otimes \sigma)V^{\dag} \zeta] =&\sum_{n=0}^{\infty} p_n \Tr[V~ (|0\>\<0| \otimes |\phi_n\>\<\phi_n|)~ V^{\dag}~~ \zeta]\\
\leq& \sum_{n=0}^\infty p_n\Tr[V(I\otimes|\phi_n\>\<\phi_n|)V^{\dag}\zeta]=
\sum_{n}^{\infty} p_n \Tr[T_{nn} \zeta]\label{bnd}= 2,
\end{split}
\end{equation}
and since $\Xi$ achieves the
bound (\ref{bnd}), it is optimal. Moreover $\Xi$ is the
unique optimal seed. In fact, the equality in (\ref{bnd}) is
achieved if and only if $\Tr[V (|0\>\<0| \otimes
|\phi_n\>\<\phi_n|)V^{\dag} \zeta] = \Tr[V(I\otimes |\phi_n\>\<\phi_n|)V^{\dag}
\zeta]$ for any $n$: by expanding the identity on the Fock basis, the positivity of $\zeta$ implies $\<m|\<\phi_n|V^\dag \zeta V|m\>|\phi_n\>=0$ for any $m \not =0$. Hence the unique nonzero diagonal elements of $\zeta$ are on the vectors $V|0\>|\phi_n\>$. 
On the other hand, the positivity of $\zeta$ along with the normalization constraint 
$\Tr[T_{mn} \zeta]=0 \quad \forall m \not =n$ imply that all the off diagonal elements of $\zeta$ 
are zero.  Hence $\zeta=2V\sum_{n=1}^\infty (|0\>\< 0|\otimes|\phi_n\>\<\phi_n|)V^\dag
=2V(|0\>\< 0|\otimes I)V^\dag=\Xi$. The fact that $\Xi$ is the unique optimal seed ensures that it
is also extremal, otherwise there would be two different seeds which are equally optimal. Notice that
$\Xi$ is extremal also according to our characterization (\ref{iff}).


\begin{thebibliography}{50}
\bibitem{note} Notice that in infinite dimensions the POVM density, whence $\Xi$, can become
unbounded, or not even anymore an operator, and it can be treated rigorously in the framework of forms.
\end{thebibliography}
\end{document}